\documentclass[10pt]{article}

\usepackage{epsfig}
\usepackage{a4wide}
\usepackage{latexsym}
\usepackage{amssymb}
\usepackage{amsmath}

\begin{document}

\def\O{{\cal O}}
\def\N{{\cal N}}
\def\>t{>_{\scriptscriptstyle{\rm T}}}
\def\enu{\epsilon_\nu}
\def\pint{\int{\d^3p\over(2\pi)^3}}
\def\gint{\int[\D g]\P[g]}
\def\hxi{\hat x_i}
\def\hatx{{\bf \hat x}}
\def\d{{\rm d}}
\def\e{{\bf e}}
\def\x{{\bf x}}
\def\X{{\bf X}}
\def\0x{\x^\smalze}
\def\sperpx{{x_\perp}}
\def\sperpk{{k_\perp}}
\def\sbperpk{{{\bf k}_\perp}}
\def\sbperpx{{{\bf x}_\perp}}
\def\perpx{{x_{\rm S}}}
\def\perpk{{k_{\rm S}}}
\def\bperpk{{{\bf k}_{\rm S}}}
\def\bperpx{{{\bf x}_{\rm S}}}
\def\r{{\bf r}}
\def\q{{\bf q}}
\def\zr{{\bf z}}
\def\R{{\bf R}}
\def\A{{\bf A}}
\def\v{{\bf v}}
\def\u{{\bf u}}
\def\w{{\bf w}}
\def\U{{\bf U}}
\def\cm{{\rm cm}}
\def\l{{\bf l}}
\def\sec{{\rm sec}}
\def\Ckol{C_{Kol}}
\def\flux{\bar\epsilon}
\def\zq{{\zeta_q}}
\def\b{b_{kpq}}
\def\bun{b^{\scriptscriptstyle (1)}_{kpq}}
\def\bdu{b^{\scriptscriptstyle (2)}_{kpq}}
\def\z0q{{\zeta^{\scriptscriptstyle{0}}_q}}
\def\smalS{{\scriptscriptstyle {\rm S}}}
\def\smalze{{\scriptscriptstyle (0)}}
\def\smalel{{\scriptscriptstyle (l)}}
\def\smalun{{\scriptscriptstyle (1)}}
\def\smaldu{{\scriptscriptstyle (2)}}
\def\smaltr{{\scriptscriptstyle (3)}}
\def\smalL{{\scriptscriptstyle{\rm L}}}
\def\smalI{{\scriptscriptstyle {\rm I}}}
\def\smalD{{\scriptscriptstyle{\rm D}}}
\def\smal1n{{\scriptscriptstyle (1,n)}}
\def\smaln{{\scriptscriptstyle (n)}}
\def\smalA{{\scriptscriptstyle {\rm A}}}
\def\shell{{\tt S}}
\def\ball{{\tt B}}
\def\nav{\bar N}
\def\micron{\mu{\rm m}}
\font\brm=cmr10 at 24truept
\font\bfm=cmbx10 at 15truept

\baselineskip 0.7cm

\centerline{\brm Forces and torques on non-spherical}
\centerline{\brm particles in viscous shear flows}
\vskip 20pt
\centerline{Piero Olla}
\vskip 5pt
\centerline{ISIAtA-CNR}
\centerline{Universit\'a di Lecce}
\centerline{73100 Lecce Italy}
\vskip 20pt

\centerline{\bf Abstract}
\vskip 5pt
The behavior of a non-spherical particle in a viscous, plane channel flow is studied by means of 
a combination of analytical technique and geometrical reasoning. An efficient implementation 
of Lamb's general solution is adopted, allowing rapid calculation, in the perturbative 
quasi-spherical regime, of forces and torques coming from the particle interaction with 
the flow gradients and from the presence of the walls. Particular attention is given to 
the role of symmetries in 
the problem. A classification of the effects of departures from spherical shape on the
motion of a particle in various kinds of shear flow is provided. Additional applications to
problems involving particles with streaming boundaries are presented.

\vskip 15pt
\noindent PACS numbers: 47.15.Pn, 47.55.Kf, 83.10.Lk,
\vfill\eject

\centerline{\bf I. Introduction}
\vskip 5pt
The motion of a spheroidal particle in a viscous shear flow is one of the classical problems
of fluid mechanics, with important applications in diverse fields such as deposition and 
sedimentation processes \cite{russel77}, hemodynamics \cite{kraus96,olla99a}, swimming of 
micro-organisms \cite{felderhof94a,felderhof94b,stone96} and the study of diffusion 
in suspensions \cite{leighton87,acrivos92}. It is well 
known that the simplest suspension model, in which the particles are rigid and spherical, 
and the flow is purely viscous, does not account for important physical processes such as 
the transverse migration of particles with respect to the flow lines. This is basically
a consequence of the symmetry properties of the Stokes equation which describes the dynamics
of a fluid in creeping flow conditions.

Different physical mechanisms leading to the symmetry breaking necessary for transverse 
migration have been analyzed in the literature \cite{leal80}. Besides the effect of 
non-spherical shapes, 
the role of inertia was considered, both in the presence of walls bounding the fluid 
\cite{ho74,vasseur76} and in infinite space situations \cite{saffman65} (see 
\cite{mclaughlin93} for recent references), while, Chan \& Leal \cite{chan79}, analyzed 
the role of non-Newtonian corrections to the constitutive law of the suspending fluid. 

The velocity disturbance produced around a non-spherical particle is very difficult to 
calculate, and, with a few exceptions such as the ellipsoid in a linear shear
(see e.g. \cite{jeffery22}), no analytical solutions exist. For shapes close to 
spherical, however, perturbation theory becomes possible and analysis of the modification to
the particle dynamics produced by different kinds of deviation from sphericity and of
external flow, can be carried on. 

An approach to the establishment of a perturbation theory is the one used e.g. in \cite{chan79}
and in \cite{stone96}, based on the use of the reciprocal theorem \cite{lorentz06,happel}. One
of the advantages of this approach consists of the fact that no information on the perturbation
produced by the particle in the given fluid flow is necessary. Rather, the simpler problem of the
translation (or rotation) of an identically shaped particle in a quiescent fluid must be
considered.

Direct perturbation expansion of Stokes equation, using Lamb's general representation 
\cite{happel}, however, has the advantage of highlighting the symmetry structure of
the fluid perturbation and its relation to that of the particle shape. The problem is
that a grater amount of information seems to be necessary in order to determine the forces
and torques necessary to obtain the particle law of motion.

It turns out that it is possible to cast this representation into a form requiring an 
amount of perturbed velocity fields calculation, equivalent to that of the reciprocal 
theorem approach (see also \cite{nadim91}). In the following sections, it will be shown 
how this can be done
using an approach that is basically an adaptation of standard quantum mechanical matrix 
techniques. In this approach, the forces, torques and lift velocities can be obtained,
in the spherical case, as components of the external flow velocity on the particle boundary
and of possible streaming velocities of the particle surface. In the non-spherical case, 
calculating such forces and torques results in the computation of matrix elements 
involving these basis functions and the unperturbed velocity field (plus streaming), 
but not the velocity perturbation by the particle.

In the next two sections Lamb's representation will be recalled and the formalism 
for the calculation of forces and torques will be derived. In section IV, some applications
to problems involving spherical particles with moving boundaries will be described;
in particular the calculation of a bound for the swimming efficiency of micro-organisms,
which is lower than the one obtained in \cite{stone96}, will be given.  
In section V, the behavior of a spheroidal particle in a plane channel flow will be considered,
separating the various contributions from interaction with the walls and with the flow 
gradients. Various geometrical proofs on the functional form of forces and torques and
on their relation to symmetry properties of the particles will be presented in the
general strongly non-spherical case. Section VI will be devoted to the conclusions.
Details of the calculations of forces and torques, and of the effect of the walls will be 
presented in the appendices.
\vskip 10pt

\centerline{\bf II. Review of Lamb's representation}
\vskip 5pt
In a perturbative analysis around a spherically symmetric situation, it is convenient to
expand the velocity field in terms of vector spherical harmonics ${\bf Y}^\mu_{lm}(\hxi)$,
$\mu=$s,l,m:
$$
\v(\x)=\sum_{lm}[v^{\rm s}_{lm}(x){\bf Y}^{\rm s}_{lm}(\hxi)+
	 v^{\rm e}_{lm}(x){\bf Y}^{\rm e}_{lm}(\hxi)+
	 v^{\rm m}_{lm}(x){\bf Y}^{\rm m}_{lm}(\hxi)], 
\eqno(2.1)
$$
where $\hxi=x_i/x$ are the cosines of $\x$ in an appropriate reference frame and the
superscripts $\{ {\rm sem}\}$, standing for scalar, electric and magnetic, come from
the origin of this basis as a tool in the study of electromagnetic waves (see \cite{landau4})
In terms of standard spherical harmonics the functions ${\bf Y}^\mu_{lm}$
are defined as
$$
{\bf Y}^{\rm s}_{lm}(\hxi)=\hatx Y_{lm}(\hxi)\qquad
{\bf Y}^{\rm e}_{lm}(\hxi)=\frac{x\nabla Y_{lm}(\hxi)}{\sqrt{l(l+1)}}\qquad
{\bf Y}^{\rm m}_{lm}(\hxi)=\frac{[\x\times\nabla]Y_{lm}(\hxi)}{\sqrt{l(l+1)}},
\eqno(2.2)
$$
where $\hat x=x^{-1}\x$. It is natural to adopt a standard bra-ket quantum mechanical
notation for the calculation of components and matrix elements: 
$|\mu l m\rangle\equiv{\bf Y}^\mu_{lm}$, $\langle\mu l m|\equiv{\bf Y}^{\mu *}_{lm}$ and
also:
$$
\langle \mu l m|{\bf v}\rangle=\int\d\Omega_x{\bf Y}^{\mu *}_{lm}(\hxi)\cdot{\bf v}(\x)
\quad{\rm and}\quad
\langle \mu l m|f|{\bf v}\rangle=\int\d\Omega_x{\bf Y}^{\mu *}_{lm}(\hxi)\cdot{\bf v}(\x)f(\x)
\eqno(2.3)
$$
Here, $\d\Omega_x$ is the solid angle differential around $\x$ and the left hand side still
depends on the radial coordinate $x$.
With the definition provided by Eqn. (2.2) (which differs from the one adopted in \cite{olla97a}, 
the vector spherical harmonics result 
normalized to one: $\langle\mu'l'm'|\mu lm\rangle=1$. Hence, one has the standard expression
for velocity components $v^\mu_{lm}(x)=\langle \mu l m|{\bf v}\rangle$. These components can 
be expressed in terms of plain spherical harmonics integrals (see Appendix A):
$$
\begin{cases}
v^{\rm s}_{lm}(R)=\int\d\Omega_x Y_{lm}^*(\hxi)\hat\x\cdot\v(R\hat\x)\\
v^{\rm e}_{lm}(R)=(l(l+1))^{-\frac{1}{2}}\int\d\Omega_x Y_{lm}^*(\hxi)
                 [3\hat\x\cdot \v(R\hat\x)-\nabla\cdot (x\v(R\hat\x))]_{x=R}\\
v^{\rm e}_{lm}(R)=-(l(l+1))^{-\frac{1}{2}}\int\d\Omega_x Y_{lm}^*(\hxi)
                 \nabla\cdot [\v(R\hat\x)\times\x]_{x=R}\\
\end{cases}
\eqno(2.4)
$$
At stationarity, an incompressible fluid in creeping flow conditions, obeys the
Stokes and continuity equations:
$$
\mu\nabla^2\v=\nabla P
\quad{\rm and}\quad
\nabla\cdot\v=0,
\eqno(2.5)
$$
where $\mu$ and $P$ are the dynamic viscosity and the pressure.
In terms of vector spherical harmonics, Lamb's representation for the solution to eqns. (2.3-4)
can be cast into the form:
$$
\begin{cases}
v^{\rm s}_{lm}=a_{lm}x^{-l}+b_{lm}x^{-2-l},\\
v^{\rm e}_{lm}=\frac{2-l}{\sqrt{l(l+1)}}a_{lm}x^{-l}-\sqrt{\frac{l}{l+1}}b_{lm}x^{-2-l},\\
v^{\rm m}_{lm}=c_{lm}x^{-1-l};\\
\end{cases}
\eqno(2.6a)
$$
in an external domain, and:
$$
\begin{cases}
v^{\rm s}_{lm}=a_{lm}x^{l+1}+b_{lm}x^{l-1},\\
v^{\rm e}_{lm}=\frac{l+3}{\sqrt{l(l+1)}}a_{lm}x^{l+1}+\sqrt{\frac{l+1}{l}}b_{lm}x^{l-1},\\
v^{\rm m}_{lm}=c_{lm}x^l.\\
\end{cases}
\eqno(2.6b)
$$
in an internal domain. One finds also for the pressure field, respectively
in an external and internal domain:
$$
P=\mu\sum_{lm}\frac{4l-2}{l+1}a_{lm}x^{-l-1}Y_{lm}\quad{\rm and}\quad
P=\mu\sum_{lm}\frac{4l+6}{l}a_{lm}x^lY_{lm}.
\eqno(2.7)
$$
and for the vorticity ${\boldsymbol{\omega}}=\nabla\times\v$, in an external domain:
$$
\boldsymbol{\omega}=\frac{1}{x}\sum_{lm}\Big(-\sqrt{l(l+1)}c_{lm}x^{-1-l}{\bf Y}^{\rm s}_{lm}
+lc_{lm}x^{-1-l}{\bf Y}^{\rm e}_{lm}
-\frac{4l-2}{\sqrt{l(l+1)}}a_{lm}x^{-l}{\bf Y}^{\rm m}_{lm}\Big);
\eqno(2.8a)
$$
and in the internal domain:
$$
\boldsymbol{\omega}=\frac{1}{x}\sum_{lm}\Big(-\sqrt{l(l+1)}c_{lm}x^l{\bf Y}^{\rm s}_{lm}
-(l+1)c_{lm}x^l{\bf Y}^{\rm e}_{lm}
+\frac{4l+6}{\sqrt{l(l+1)}}a_{lm}x^{l+1}{\bf Y}^{\rm m}_{lm}\Big).
\eqno(2.8b)
$$
The velocity field is thus partitioned into contributions from the pressure (the $a_{lm}$ terms),
and into purely potential flow (the $b_{lm}$ terms) and pressure free, vortical
contributions (the $c_{lm}$ terms). 

\vskip 5pt
A sequence of formulae of practical use can be derived. Expressing the components of the velocity 
field in terms of velocity boundary conditions at a reference distance $x=R$: 
$\v(R\hat\x)=\U(\hat\x)$, one has, in an external domain:
$$
\begin{cases}
v^{\rm s}_{lm}=
\frac{1}{2}\Big[\Big(l+(2-l)\Big(\frac{R}{x}\Big)^2\Big)U^{\rm s}_{lm}
+\sqrt{l(l+1)}\Big(l-\Big(\frac{R}{x}\Big)^2\Big)U^{\rm e}_{lm}\Big]\Big(\frac{R}{x}\Big)^l\\
v^{\rm e}_{lm}=
\frac{1}{2}\Big[(2-l)\sqrt{\frac{l}{l+1}}\Big(1-\Big(\frac{R}{x}\Big)^2\Big)U^{\rm s}_{lm}
+\Big(2-l+l\Big(\frac{R}{x}\Big)^2\Big)U^{\rm e}_{lm}\Big]\Big(\frac{R}{x}\Big)^l\\
v^{\rm m}_{lm}=U^{\rm m}_{lm}\Big(\frac{R}{x}\Big)^{l+1}\\
\end{cases}
\eqno(2.9a)
$$
while, in an internal domain:
$$
\begin{cases}
v^{\rm s}_{lm}=
\frac{1}{2}\Big[\Big(-(l+1)+(l+3)\Big(\frac{R}{x}\Big)^2\Big)U^{\rm s}_{lm}
+\sqrt{l(l+1)}\Big(1-\Big(\frac{R}{x}\Big)^2\Big)U^{\rm e}_{lm}\Big]\Big(\frac{x}{R}\Big)^{l+1}\\
v^{\rm e}_{lm}=
\frac{1}{2}\Big[(l+3)\sqrt{\frac{l+1}{l}}\Big(-1+\Big(\frac{R}{x}\Big)^2\Big)U^{\rm s}_{lm}
+\Big(l+3-(l+1)\Big(\frac{R}{x}\Big)^2\Big)U^{\rm e}_{lm}\Big]\Big(\frac{x}{R}\Big)^{l+1}\\
v^{\rm m}_{lm}=U^{\rm m}_{lm}\Big(\frac{x}{R}\Big)^l\\
\end{cases}
\eqno(2.9b)
$$
Similarly, it is possible to express the force on a surface bounding the fluid in terms of
the velocity components on that surface.
The force density ${\bf f}$ exerted by the fluid on a spherical surface element at $x=R$, is 
given by the normal component of the stress tensor $S_{ij}=-P\delta_{ij}+\sigma_{ij}$: 
${\bf f}=\pm\hat\x\cdot{\bf S}$,
where $\sigma_{ij}=\mu(\partial_iv_j+\partial_jv_i)$ is the viscous stress and the $\pm$ signs
refer to the case of the surface bounding the fluid, respectively, from the inside and from the
outside. 
Using Eqns. (2.7) and (2.9a-b), one finds, in the case of an external domain (see also 
\cite{brenner64}):
$$
\begin{cases}
f^{\rm s}_{lm}=\frac{\mu}{R}
\Big[-\frac{2l^2+3l+4}{l+1}U^{\rm s}_{lm}+3\sqrt{\frac{l}{l+1}}U^{\rm e}_{lm}\Big]\\
f^{\rm e}_{lm}=\frac{\mu}{R}
\Big[3\sqrt{\frac{l}{l+1}}U^{\rm s}_{lm}-(2l+1)U^{\rm e}_{lm}\Big]\\
f^{\rm m}_{lm}=-\frac{\mu}{R}(l+2)U^{\rm m}_{lm}\\
\end{cases}
\eqno(2.10a)
$$
and 
$$
\begin{cases}
f^{\rm s}_{lm}=\frac{\mu}{R}
\Big[-\frac{2l^2+l+3}{l}U^{\rm s}_{lm}+3\sqrt{\frac{l+1}{l}}U^{\rm e}_{lm}\Big]\\
f^{\rm e}_{lm}=\frac{\mu}{R}
\Big[3\sqrt{\frac{l+1}{l}}U^{\rm s}_{lm}-(2l+1)U^{\rm e}_{lm}\Big]\\
f^{\rm m}_{lm}=-\frac{\mu}{R}(l-1)U^{\rm m}_{lm}\\
\end{cases}
\eqno(2.10b)
$$
in that of an internal domain. Inverting Eqns. (2.10b) and substituting into Eqns. (2.9a-b),
it is possible to obtain expressions
for the velocity field produced by a distribution of forces on the surface $x=R$.
\vskip 20pt

\centerline{\bf III Representation of forces and torques on particles}
\vskip 5pt
The force ${\bf F}$ exerted on a particle by the fluid is obtained integrating the 
expression for the force density 
provided by Eqns. (2.10a-b) on a spherical surface surrounding that particle. As it is
well known \cite{happel}, the only term contributing to the force is the one coming from
the $a_{1m}$ components of the velocity perturbation $[$see Eqn. (2.6a)$]$:
$$
F_1=2\mu\sqrt{6\pi}Re a_{11};\qquad F_2=-2\mu\sqrt{6\pi}Im a_{11};
\qquad F_3=-2\mu\sqrt{3\pi}a_{10},
\eqno(3.1)
$$
Now, the coefficients $a_{1m}$ can be expressed in bra-ket notation as:
$a_{11}=\frac{x}{2}\{\langle{\rm s}11|+\sqrt{2}\langle{\rm e}11|\}|\v\rangle$ and similar 
for $m\ne 1$. Thus, the force on the particle can be obtained calculating the component
of the perturbed velocity on adequate basis functions, which are combinations of ''scalar''
and ''electric'' spherical harmonics, and which have the expression:
$$
|F_1\rangle=\mu x[-3/2,0,0];\qquad |F_2\rangle=\mu x[0,-3/2,0];
\qquad
|F_3\rangle=\mu x[0,0,-3/2];
\eqno(3.2)
$$
The torque ${\bf T}$ can be calculated in a similar way, starting from the expression 
$T_i=\int\d\Omega_x\epsilon_{ijk}x_jf_kx^2$ $=\int\d\Omega_x\epsilon_{ijk}x_jx_l\sigma_{lk}$
and this leads to the result that now the ''magnetic'' $l=1$ components of the velocity
perturbation are to be taken in exam \cite{happel}:
$$
T_1=-4\sqrt{3\pi}\mu Re c_{11};\qquad T_2=4\sqrt{3\pi}\mu Im c_{11};\qquad
T_3=2\sqrt{6\pi}\mu c_{10}
\eqno(3.3)
$$
In a way analogous to the case of the force, the torque can then be calculated extracting the 
appropriate components of the velocity field. The associated basis functions are proportional 
to ''magnetic'' harmonics with $l=1$:
$$
|T_1\rangle=3\mu x[0,x_3,-x_2];\qquad |T_2\rangle=3\mu x[x_3,0,-x_1];
\qquad
|T_3\rangle=3\mu x[x_2,-x_1,0].
\eqno(3.4)
$$
In principle, Eqns. (3.2) and (3.4) require knowledge of the velocity perturbation around
the particle to calculate force and torque components, however, for particle shapes close
to spherical this is not the case. In the case of a radius $R$ spherical particle, no-slip 
boundary conditions on the particle surface imply:
$$
\hat\U(\hat\x)=\bar\U(\hat\x)+\U(\hat\x)
\eqno(3.5)
$$
where $\bar\U$ and $\U$ are the value at $x=R$ of the external flow $\bar\v$ and of the velocity 
perturbation $\v$,
as seen in the reference frame translating with the particle, and, apart of the components 
$\hat U^{\rm m}_{1m}$, representing a bulk rotation, $\hat\U$ is non-zero only if there is 
streaming of the particle surface (or boundary conditions which allow for the presence of slip).
Hence, the force and the torque on this particle will be given by the expressions:
$$
{\bf F}=\langle{\bf F}|\{\hat\U-\bar\U\}\rangle
\quad{\rm and}\quad
{\bf T}=\langle{\bf T}|\{\hat\U-\bar\U\}\rangle
\eqno(3.6)
$$
where $\langle{\bf F}|$ and $\langle{\bf T}|$ are the vectors whose components are the ''bras''
$\langle F_i|$ and $\langle T_i|$. (Clearly, in the absence of torque, $\hat U^{\rm m}_{1m}=
\bar U^{\rm m}_{1m}$).  Thus, trivially, no information on the velocity perturbation
produced by the particle in the fluid is necessary to calculate forces and torques on the
particle. This, independently of possible complication in the form of the surface streaming 
$\hat\U$, which makes this approach useful in problems of micro-organim swimming and other 
situations involving objects with moving boundaries.

\vskip 5pt
In the case of a spheroidal particle, the velocity disturbance $\v$ can be expressed as a
perturbative series $\v=\v^\smalze+\v^\smalun+...$, where $\v^\smalze$ is the value of $\v$ 
in the spherical case, given the same external flow conditions. It is standard practice to 
parametrize the deviation from spherical
shape in terms of the quantity: $\Delta X(\hat\x)=X(\hat\x)-R$, where $x=X(\hat\x)$ is the
equation for the particle surface and the particle center is fixed imposing the condition
that $X$ does not have $l=1$ components. The $\O(\Delta X/R)$ contribution to $\v$ will then be
in the form:
$$
\v^\smalun=-\Delta X\partial_x(\bar\v+\v^\smalze)_{x=R}\equiv
\Delta X\U'
\eqno(3.7)
$$
plus contribution coming from possible $\O(\Delta X/R)$ corrections to $\hat\U$. Using Eqns. 
(2.9a-b) and (3.5), the term $\U'$ can be expressed in terms of the values
$\bar\U$ and $\hat\U$ of the external shear and of the streaming velocity on the $x=R$ surface:
$$
\begin{cases}
U_{lm}^{\prime\,\rm s}=0\\
U_{lm}^{\prime\,\rm e}=\frac{2l+1}{R}\Big(\frac{3\bar U^{\rm s}_{lm}}{\sqrt{l(l+1)}}-
2\bar U^{\rm e}_{lm}\Big)+2l\hat U^{\rm e}_{lm}\\
U_{lm}^{\prime\,\rm m}=-\frac{2l+1}{R}\bar U^{\rm m}_{lm}+\frac{l+1}{R}\hat U^{\rm m}_{lm}\\
\end{cases}
\eqno(3.8)
$$
From here, the  $\O(\Delta X/R)$ contributions to force and torque can be expressed in a form 
which actually does not require information on the detailed structure of the velocity disturbance;
explicitating the $\O(\Delta X/R)$ corrections to $\hat\U$:
$$
{\bf F}^\smalun=\langle{\bf F}|\hat\U^\smalun\rangle+
\sum_{\mu lm}\langle{\bf F}|\Delta X|\mu l m\rangle U_{lm}^{\prime\,\mu}
\eqno(3.9)
$$
and
$$
{\bf T}^\smalun=\langle{\bf T}|\hat\U^\smalun\rangle+
\sum_{\mu lm}\langle{\bf T}|\Delta X|\mu l m\rangle U_{lm}^{\prime\,\mu}.
\eqno(3.10)
$$
In fact, the matrix elements in Eqns. (3.9-10) require the calculation of integrals, which are
basically equivalent to the ones needed in the reciprocal theorem approach \cite{happel},
say
$
\int_A\d{\bf A}\cdot{\boldsymbol{\sigma}}\cdot(\hat\U-\bar\U),
$
where $A$ is the particle surface defined by $x=X(\hat\x)$, and ${\boldsymbol{\sigma}}$ is the 
viscous stress
generated around the particle when this is either translating or rotating in a quiescent fluid.
\vfill\eject

\centerline{\bf IV. Spherical particles with generic boundary conditions at the surface}
\vskip 5pt
As a first application of the formalism illustrated in the previous section, the drag 
force and torque acting on a rigid, radius $R$ spherical particle, carrying on 
respectively a translational and a rotational motion, can readily be calculated. In the case
of a translation with velocity $U$, one finds from Eqn. (3.2): 
$$
F_i=\langle F_i|{\bf U}\rangle=-(3/2)\mu RU\int\d\Omega_x=-6\pi\mu RU, 
\eqno(4.1)
$$
which is the standard Stokes 
formula for the drag. In the rotation case, take the angular velocity 
directed along $x_3$, 
with magnitude $\omega$. The velocity at the particle surface is then $\U=
\omega[-x_2,x_1,0]$ and the drag will be:
$$
T_3=\langle T_3|\U\rangle=-3\omega\mu R^3\int\d z\d\phi(1-z^2)=-8\pi R^3\mu\omega.
\eqno(4.2)
$$
As a less obvious example, consider the velocity of a spherical
particle at the centre of a quadratic shear flow: $\bar\v=\kappa x^2_2\e_3$. From Eqn. (3.2), 
the force is 
$F_3=-\langle F_3|\bar\U\rangle=\mu \kappa R^32\pi\int_{-1}^1\d z z^2=2\mu\pi\kappa R^3$.
Using the expression for Stokes drag, one obtains that the particle will move with a velocity 
$$
u=(6\pi\mu R)^{-1}F_3=\frac{\kappa R^2}{3}\e_3.
\eqno(4.3)
$$
The same result could have obtained using Faxen's equation for the force \cite{happel}, 
and, in a certain sense, Eqn. (3.6) can be seen as an integral form of Faxen's
result, with the bonus that boundary conditions different from no-slip can be taken into 
account explicitly, and that perturbation theory becomes possible.
\vskip 5pt 
A typical situation with moving boundaries is that of a swimming micro-organism. It is then 
possible to obtain an upper bound for the swimming efficiency, more precise than the one derived 
in \cite{stone96}. A definition for the swimming efficiency of a radius $R$ spherical 
micro-organism is the following one, due to Lighthill \cite{childress}:
$$
\eta_L=\frac{6\pi\mu Ru^2}{W}
\eqno(4.4)
$$
where $6\pi\mu Ru^2$, from Eqn. (4.1), is the power necessary to set an identically shaped rigid 
particle in motion at speed $u$, and $W=\int\d A {\bf f}\cdot\U$ is the work per unit time 
actually carried on in swimming.
From Eqns. (2.10a) and (3.5), in the case of a particle which does not change shape 
while swimming, this work is equal to:
$$
W=\mu R\sum_{lm}\Big[(2l+1)\Big(\frac{-3\bar U^{\rm s}_{lm}}{\sqrt{l(l+1)}}+2\bar U^{\rm e}_{lm}
-\hat U^{\rm e}_{lm}\big)\hat U^{\rm e*}_{lm}+\Big((2l+1)\bar U^{\rm m}_{lm}
-(l+2)\hat U^{\rm m}_{lm}\Big)\hat U^{\rm m*}_{lm}\Big]
\eqno(4.5)
$$
where $\hat\U$ is the streaming velocity associated with the swimming stroke and $\bar\U=-\u$.
If $\u$ is along $x_3$, it is easy to show that, necessarily:
$$
u=-\frac{\langle F_3|{\rm e}10\rangle}{6\pi\mu R}\hat U^{\rm e}_{10}
=-\sqrt{6\pi}\hat U^{\rm e}_{10}
\eqno(4.6)
$$
where the expression for the vector spherical harmonic 
${\bf Y}^{\rm e}_{10}=\sqrt{\frac{3}{8\pi}}\Big([0,0,1]-\frac{x_3\x}{x^2}\Big)$ 
has been used $[$see Eqn. (2.2)$]$. In other words, only the $|{\rm e}10\rangle$ component of 
$\hat\U$ contributes to the force on the particle (there are no ''s'' components, in
any case, due to absence of particle changes of shape). From Eqn. (4.5) and using also the 
expression 
${\bf Y}^{\rm s}_{10}=\sqrt{\frac{3}{4\pi}}\frac{x_3\x}{x^2}$, one obtains for the work performed
in swimming:
$$
W\ge 2\mu R\Big(-\frac{3\bar U^{\rm s}_{10}}{\sqrt{2}}+2\bar U^{\rm e}_{10}
-\hat U^{\rm e}_{10}\Big)\hat U^{\rm e*}_{10}=12\pi\mu R u^2
\eqno(4.7)
$$
which sets the maximum swimming efficiency to $\eta_L=\frac{1}{2}$, in the case the velocity 
associated with the ''swimmimg stroke'' is a steady $|{\rm e}10\rangle$ circulation on the 
particle surface. This upper 
bound is stronger than the condition $\eta_L<\frac{3}{4}$ presented in \cite{stone96}, due to the
neglect in Eqn. (11) of that reference of the contribution to dissipation from vorticity, which 
from Eqn. (2.8a) is given by:
$$
H=\int\d^3 x\omega^2\ge\frac{\mu R}{2}|U^{\rm e}_{10}|^2
\eqno(4.8)
$$
\vskip 5pt
As a last exercise, it is interesting to generalize the result provided by Eqn. (3.6) for the force
on a spherical particle, to the case of an infinite surface tension fluid droplet. This 
clearly parallels the generalization of Faxen's equation, obtained in \cite{rallison78} using 
the reciprocal theorem.  The boundary conditions for this problem are continuity
of tangential stresses across the interface and absence of deformations from spherical shape, 
which means $\hat U^{\rm s}_{lm}=0$. Focusing on the $x_3$ component of the force, it will be 
necessary to know only the $|{\rm e}10\rangle$ component of $\hat\U$:
$$
F_3=-\langle F_3|\bar\U\rangle
+\langle F_3|{\rm e}10\rangle\hat U^{\rm e}_{10}
\eqno(4.9)
$$
In order to calculate the velocity component 
$\hat U^{\rm e}_{10}$, the necessary boundary conditions are:
$$
U^{\rm e}_{10}=\hat U^{\rm e}_{10}-\bar U^{\rm e}_{10},
\qquad
U^{\rm s}_{10}=-\bar U^{\rm s}_{10}
\quad{\rm and}\quad
f^{\rm e}_{10}=\bar f^{\rm e}_{10}+\hat f^{\rm e}_{10} 
\eqno(4.10)
$$
where $f^{\rm e}_{10}$ is given by Eqn. (2.10a) while $\bar f^{\rm e}_{10}$ and 
$\hat f^{\rm e}_{10}$ are given by Eqn. (2.10b). One obtains promptly:
$$
\hat U^{\rm e}_{10}=
\frac{\sqrt{2}}{1+\hat\mu/\mu}\Big(\sqrt{2}\bar U^{\rm e}_{10}-\frac{3}{2}\bar U^{\rm s}_{10}\Big)
\eqno(4.11)
$$
where $\hat\mu$ is the droplet fluid viscosity. Substituting Eqns. (4.11) and (3.2), 
into Eqn. (4.9) and using the expressions for ${\bf Y}^{\rm s,e}_{10}$ previously given, 
one finds an expression for the force, again in the form provided by Eqn. (3.6):
$F_i(\hat\mu/\mu)=
-\langle F_i(\bar\mu/\mu)|\hat\U\rangle$, where now:
$$
|F_i(\lambda)\rangle=-\frac{3}{2}\mu R(1+\lambda)^{-1}((\lambda-1)\e_i+5\hxi\hat\x)
\eqno(4.12)
$$
Clearly, Eqn. (3.2) is recovered in the limit $\lambda=\hat\mu/\mu\to\infty$. In the 
quadratic shear case, discussed in Eqn. (4.3), the force on the droplet will be: 
$F_3=\frac{2\pi\mu\kappa R^3}{1+\mu/\hat\mu}$, which is zero in the limit $\hat\mu/\mu\to 0$,
typical of a spherical bubble.
\vskip 20pt

\centerline{\bf V. Spheroidal rigid particle in plane channel flow}
\vskip 5pt
\noindent{\bf A. Geometrical approach}
\vskip 5pt
\noindent The presence of solid surfaces bounding the flow leads to additional forces and torques 
on particles in suspension; these forces and torques can be obtained from Eqns. (3.6) 
and (3-9-10), once the modifications produced on the external velocity field $\bar\v$ are known. 
If the distance $l$ from the wall is
large compared with the particle radius $R$, it will be possible to approximate the
modification to $\bar\v$, with the first image $\v^\smalI$ at the wall, of the velocity 
perturbation by the particle. This image is obtained 
imposing on the wall the no-slip boundary condition $\v^\smalI+\v=0$, where $\v$ is the velocity 
perturbation by the particle in the infinite domain case.

If the particle is neutrally buoyant, the velocity perturbation at the wall will be 
$\v=\O((R/l)^2)$ and similarly, one will have that $v^\smalI=\O((R/l)^2)$ at the particle, 
while stress and vorticity will be $O((R/l)^3)$. In the far field regime, the main effect 
of the wall is therefore a force, which, as it is well known, will have a transverse 
component only if the particle is non-spherical. As illustrated in Fig. 1, a lot of what 
happens can be understood in terms of the particle and the flow symmetry properties, at
least if the shear plane is also a plane of symmetry for the particle.

\begin{figure}[hbtp]\centering
\centerline{
\psfig{figure=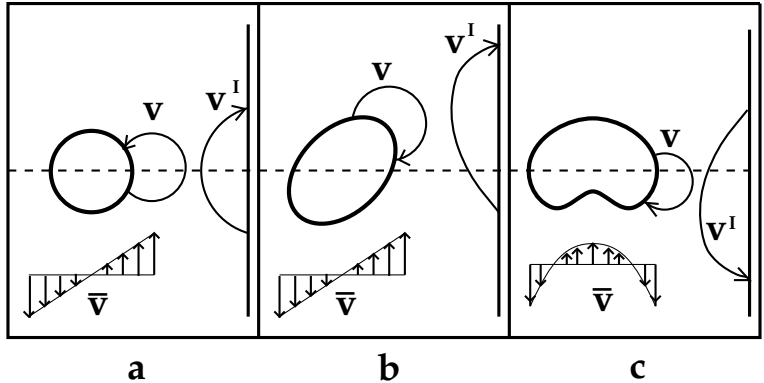,height=10.cm,angle=-90.}
}
\caption{Sketch of velocity perturbation around a particle in various shear flows, and image
velocity field produced by a wall; the velocity fields are all in the particle's reference frame.
(a) perturbation around a spherical particle: the velocity lines are symmetric around the 
perpendicular to the wall (dashed line) and no lift is present; 
(b) perturbation by an ellipsoidal particle in linear shear: as in the spherical case, the field
lines are characterized by a quadrupole symmetry; (c) the same for a particle without fore-aft 
symmetry in quadratic shear: in the near field, the velocity perturbation has an octupole symmetry.
}
\end{figure}
Definite results can be obtained in the presence of one more symmetry plane, 
identifying a fore-aft line for the particle, or if the particle has fore-aft
symmetry, if there are in total three symmetry planes.
It is clear that in the two cases $b$ and $c$, the lift will be proportional respectively to 
$\sin 2\theta$ and to $\cos\theta$, where $\theta$ is the angle of the particle symmetry axis 
with respect to the flow direction, as shown in figure. 

Transversal drifts arise also from the interaction of the particle non-spherical shape with 
the flow gradient. For instance, as shown in Fig. 2, a particle without fore-aft symmetry 
immersed in a linear shear flow will feel a force perpendicular to its symmetry axis.
\begin{figure}[hbtp]\centering
\centerline{
\psfig{figure=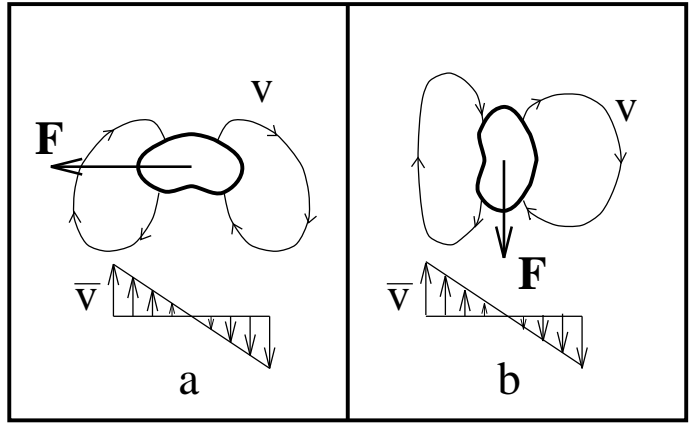,height=9.cm,angle=-90.}
}
\caption{Forces on particles lacking fore-aft symmetry.
In case (a), top velocity lines (going towards the right) do not correspond to bottom ones
(going to the left); the end result is a force in the horizontal direction. To each line
going upwards, instead, there is a symmetric one going downwards (see e.g. portions of the 
velocity lines to the extreme left and extreme right of the particle). Hence, no vertical
component in the velocity is present. Analogous reasoning shows that in case (b), fore-aft
asymmetry leads to a vertical force.} 
\end{figure}
This means an additional drift proportional to $\cos\theta$, analogous to the one
described in Fig. 1c, in the case of a non fore-aft symmetric particle interacting with a
wall, in a quadratic shear.
In the perturbative case, this force arises from the combination in Eqn. (3.9) of the $l=2$ 
harmonics in the external shear and the $l=3$ components in $\Delta X$. The same line of 
reasoning would lead to expect that an ellipsoidal particle in a quadratic shear should 
feel a force that becomes maximum when the particle axis is at $45^o$ with respect to the
flow. However, the issue is not so simple: given a flow in the form $\bar\v=v_0(a+b(x_2/R)^2)\e_3$, 
the force will be in the form ${\bf F}=a{\bf F}'+b{\bf F}''$ with ${\bf F}'$ and ${\bf F}''$
in general not parallel. However, since the angle between the two force contributions is unknown, 
it is not clear what is the sign of $F_2$, when 
$a$ is chosen in such a way to have $F_3=0$. In fact, it will turn out in the 
perturbative regime, that $F_2=0$.

Turning to the torque, similar symmetry based considerations allow to select which particle
shapes and flow regimes lead to a non-zero contribution. In the perturbative regime, Eqn. 
(3.10) and the even parity of $|T\rangle$ require that $\Delta X$ and $\bar\v$ be 
characterized both by even or odd spherical harmonics, which means, from the fact that 
$|T\rangle$ is $l=1$, that these harmonics can only be equal. As with the force, this 
statement can be generalized to the strongly non-spherical regime to the result that 
quadratic shear can act with a torque only on particles whose shape is not fore-aft 
symmetric (see Fig. 3).
\begin{figure}[hbpt]\centering
\centerline{
\psfig{figure=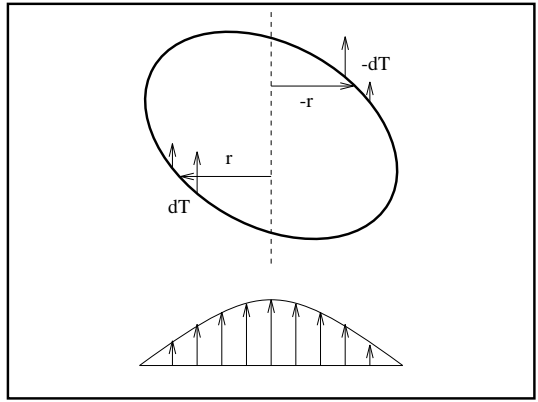,height=7.cm,angle=-90.}
}
\caption{Fore-aft symmetric particle in quadratic shear: no torques are produced.
Because of symmetry of both the particle shape and the velocity field, for any
surface element at $\r$, there is another one at $-\r$, such that the contributions
to torque are equal and opposite. }
\end{figure}
Additional informations on the dependence of the torque on the particle orientation, can be 
obtained writing the external velocity field $\bar\v$ in a reference system rotating with the
particle. To fix the ideas, take $\bar\v=\bar v(x_2)\e_3$ and, indicating with a prime the 
rotating reference frame, take $x'_1\equiv x_1$ and $x'_3$ to be along the particle symmetry
axis (hence, in Fig. 1c, the two reference frames coincide). One will have then in the
case of a linear shear:
$$
\e_3x_2=\e'_3x'_2\cos^2\theta-\e'_2x'_3\sin^2\theta
-\e'_3x'_3\sin\theta\cos\theta+\e'_2x'_2\sin\theta\cos\theta
\eqno(5.1)
$$
while, for a quadratic shear:
$$
\e_3x_2^2=\e'_2{x'_2}^2\sin\theta\cos^2\theta+\e'_2{x'_3}^2\sin^3\theta
-2\e'_3x'_2x'_3\cos^2\theta\sin\theta
$$
$$
-2\e'_2x'_2x'_3\sin^2\theta\cos\theta+\e'_3{x'_2}^2\cos^3\theta
+\e'_3{x'_3}^2\cos\theta\sin^2\theta
\eqno(5.2)
$$
Consider first the case of a fore-aft symmetric particle in a linear shear flow. As illustrated 
in Fig. 4, because of symmetry across the planes $x'_1x'_2$ and $x'_1x'_3$ the terms in 
$\sin\theta\cos\theta$ in Eqn. (5.1) do not contribute. 
\begin{figure}[hbpt]\centering
\centerline{
\psfig{figure=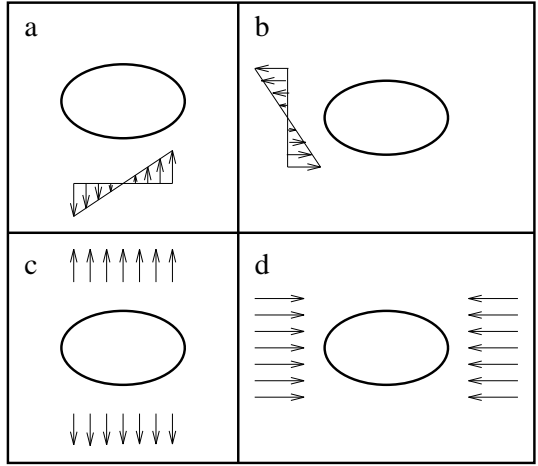,height=7.cm,angle=-90.}
}
\caption{Derivation of the functional form for the torque on a fore-at symmetric particle
in linear shear. The flow fields drawn in quadrants (a-d) correspond to the four terms, in the
decomposition of the linear shear in the primed reference frame of $\e_3x_2$ $[$see Eqn. (5.1)$]$.
Flows (c) and (d) produce boundary conditions for the velocity perturbation $\v$, which, 
because of their symmetry, do not lead to torque. 
}
\end{figure}
One recovers therefore, without having
to calculate the velocity perturbation around the particle, the expression for the torque, 
derived in \cite{jeffery22} in the case of the ellipsoid:
$$
T_1=a(1+b\cos 2\theta)
\eqno(5.3)
$$
Turning to the case of a quadratic shear and of a particle which is not fore-aft symmetric, 
it appears from analysis of Fig. 5, that only the terms on the first line of Eqn. (5.2)
contribute to torque, so that, in that case:
$$
T_1=c(1+d\cos 2\theta)\sin\theta
\eqno(5.4)
$$
\begin{figure}[hbpt]\centering
\centerline{
\psfig{figure=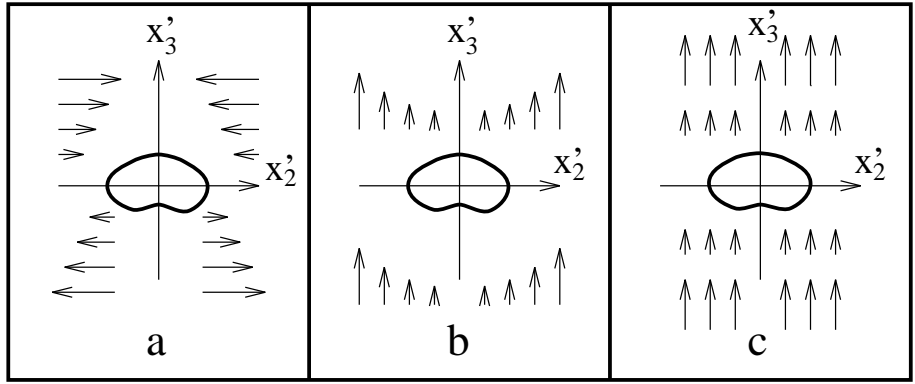,height=12.cm,angle=-90.}
}
\caption{Terms in the decomposition of a quadratic shear in terms of primed variables, which
do not contribute to torque. Flows (a-c) correspond to the terms on the second line of Eqn. 
(5.2)}
\end{figure}
Thus, quadratic shear acts in the direction to align the particle axis with the flow.
Conversely, if linear shear is the only actor in the
game and the particle is rigid, no equilibrium orientations are possible. 
(At least, this is true in the case of the rigid ellipsoid \cite{jeffery22}). 
It remains to be shown that quadratic shear plus fore-aft asymmetric shape, is actually 
the only mechanism for orientation fixing. In Fig. 6, it is shown how the last 
possibility of an odd contribution to torque from interaction between fore-aft asymmetry 
and linear shear is excluded. 
\begin{figure}[hbpt]\centering
\centerline{
\psfig{figure=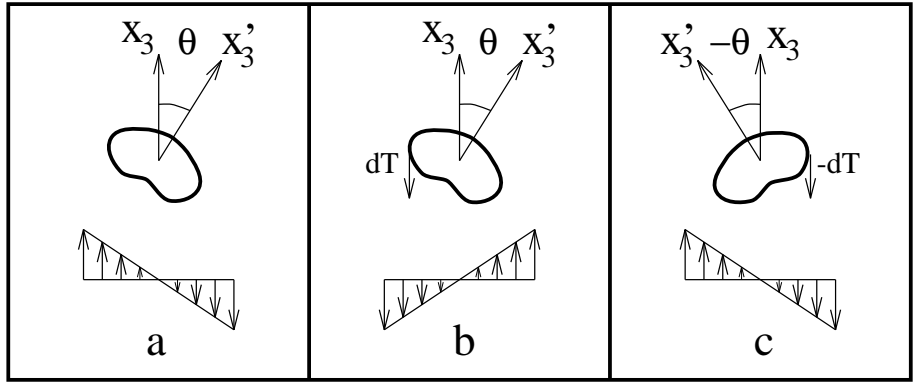,height=12.cm,angle=-90.}
}
\caption{The torque on a fore-aft asymmetric particle in a linear shear, does not
contain odd $\theta$  contributions. Case (b) is obtained from case (a) changing
sign to the velocity at all points on the particle surface. Hence the torques in (a) and
(b) are equal and opposite. Case (c) is obtained from (b) by a reflection. The velocity,
and therefore also the force, remain the same, point by point on the particle surface.
The position vector from $x_3$ to the point on the surface, however changes sign, and
then the torque is the opposite to that in case (b). Hence the torque in (a) and in (c)
are equal.}
\end{figure}

To summarize, the istantaneous transversal drift of a particle in a channel flow is due 
to three distinct mechanisms, which, for shapes which have at least two planes
of symmetry (one of which is the shear plane),  can be described as follows: 
\renewcommand{\theenumi}{\roman{enumi}}
\begin{enumerate}
\item An interaction with the walls due to the modifications in the velocity perturbation,
produced by the particle shape deviation from spherical and the linear part of the external
shear. This contribution is proportional to $\sin 2\theta$, 
with $\theta$ the angle between the direction of the flow and the 
particle symmetry axis. This drift is expected to be directed away from the nearest wall, when 
the longer particle axis is aligned with the straining direction of the flow (see Fig. 1b).
\item A second interaction with the walls, which is present when the particle shape is not 
fore-aft symmetric and the shear has a quadratic part. The resulting drift is proportional to 
$\cos\theta$, and, at least in the near field regime (see Fig. 1c), one expects that it be 
directed away from the nearest wall, when the particle's ''tip'' is in the direction of the flow.
\item A term proportional to $\cos\theta$, which is due only to the interaction of 
the particle with the gradient of $\bar\v$, and which is present only if the particle is 
not fore-aft symmetric. This drift pushes the particle towards the regions in which
$\bar\v$ is in the direction of the particle's tip (see Fig. 2a).
\end{enumerate}
Using also Eqns. (5.3-4), this means that, in a channel flow, the law of motion for a 
particle away from the walls will be in the form:
$$
\begin{cases}
\dot y=A(\theta,y,L)\sin 2\theta+B(\theta,y,L)\cos\theta\\
\dot\theta=a'(1+b\cos 2\theta)+c'(1+d\cos 2\theta)\sin\theta\\
\end{cases}
\eqno(5.5)
$$
where $y$ is the distance from a wall and $2L$ the channel width.
From previous discussion, $A$ and $B$ are even under the exchanges $\theta\to -\theta$ and 
$\theta\to\theta+n\pi/2$, $a'$ is proportional to the linear part of the shear, with  $|b|<1$; 
one has $c\ne 0$ only in the case of a fore-aft asymmetric particle in quadratic shear. 
From Eqn. (5.5) it appears that $\dot\theta(\theta)=\dot\theta(\pi-\theta)$ and at the
same time $\dot y(\theta)=-\dot y(\pi-\theta)$. Thus, unless an equilibrium orientation
$\theta_0$: $\dot\theta(\theta_0)=0$ exists for $a'\simeq c'$, $y$ will have a periodic 
time dependence and no steady drift will occur. Of course, this result is a consequence of 
Stokes flow reversibilty \cite{happel} and is therefore valid also for generic non-symmetric
particle shapes. This differs from the situation considered in 
\cite{olla99a}, where the deformability of the particle (a red cell) implied a contribution to 
the coefficient $c'$ in Eqn. (5.5) proportional to $\cos\theta$, meaning violation of the
symmetry $\dot y(\theta)=-\dot y(\pi-\theta)$ and the possibility of a mean drift also in the 
periodic regime (the region away from the channel axis corresponding to $c'$ small). In the
present case, instead, if the form of the coefficients $A$ and $B$ allows for a stable
fixed point of Eqn. (5.5) at $\theta=0,y=0$, the attraction basin of this point will be
a smaller region close to the channel axis, and not the whole space domain.

\vskip 10pt
\noindent{\bf B. Analytical results}
\vskip 5pt
\noindent The bra-ket approach of section III for the calculation of forces and torques can be 
generalized to the case of $v_\smalL$, i.e. the  part of the transverse drift due to the channel walls.
From Eqn. (2.6a), for a neutrally buoyant particle, the far field behavior of $\v$ will be in the
form: $\v=x^{-5}\x\sum_{ij}\alpha_{ij}x_ix_j$, with $\alpha_{ij}$ traceless and symmetric. Taking
the channel walls to be perpendicular to the $x_2$ axis, analysis carried on in Appendix B shows
that the particle drift in direction $x_2$ depends only on $\alpha_{22}$ and can be 
approximated 
very well superposing the images $v^\smalI$ produced separately by each wall. If the walls are 
located, with respect to the particle, at $x_2=L_r$ and $x_2=-L_l$, with $L_l,L_r>0$:
$$
v_\smalL\simeq\frac{\alpha_{22}}{4}\Big(\frac{1}{L_l^2}-\frac{1}{L_r^2}\Big)
\eqno(5.6)
$$
Following the approach of section III, one introduces a ket $|L\rangle$ 
such that: $\alpha_{22}/4=\langle L|\v\rangle$. Using Eqn. (2.9a) and the expressions for the 
spherical harmonics: $Y_{20}=\frac{1}{4}\sqrt{\frac{5}{\pi}}(3\hat x_3^2-1)$ and 
$Y_{2\pm 2}=\frac{1}{4}\sqrt{\frac{15}{2\pi}}(\hat x_1^2-\hat x_2^2\pm 2\hat x_1\hat x_2)$,
one finds: $\alpha_{22}=-\frac{1}{2}\sqrt{\frac{15}{2\pi}}\Big(U^{\rm e}_{20}
+\sqrt{\frac{3}{2}}(U^{\rm e}_{22}+U^{\rm e}_{2,-2})\Big)$ plus terms involving scalar
components $U^{\rm s}_{lm}$, which will be shown later not to contribute. From Eqn. (2.4) 
and the expression given above for $Y_{2m}$, one finds therefore:
$$
|L\rangle=\frac{15}{16\pi}[-\hat x_1\hat x_2^2,(1-\hat x_3^2)\hat x_2,-\hat x_2^2\hat x_3]
\eqno(5.7)
$$
plus the contributions associated with $U^{\rm s}_{lm}$. From here, the generalization of
Eqns. (3.6) and (3.9) can be obtained, but the first one gives identically zero, and we 
are left with:
$$
v_\smalL=\Big(\frac{1}{L_l^2}-\frac{1}{L_r^2}\Big)\sum_{\mu lm}\langle L|\Delta X|\mu lm\rangle
U^{\prime\mu}_{lm}
\eqno(5.8)
$$
and the absence of $\mu={\rm s}$ contributions in the sum $[$see Eqn. (3.8)$]$ guarantees that 
the $U^{\rm s}_{lm}$ contributions could actually be neglected in the calculation of $|L\rangle$.

The law of motion of a particle in a channel flow can then 
be determined perturbatively using the formalism provided by Eqns. (3.2), (3.4), (3.8-10) 
and (5.7-8). Indicate the transverse particle coordinate in the laboratory frame with $y$, 
and fix the origin of the axis at the left wall, so that $y\equiv L_l$; in this 
reference frame, the velocity profile is $\bar\omega y(1-y/(2L))$ where $2L=L_l+L_r$. As
before, $\{x_1,x_2,x_3\}$ is the reference frame translating with the particle center, with 
$x_3$ along the flow and $x_2$ along the direction of variation for $\bar v$; for a particle
at $y$, from Eqn. (4.3): 
$$
\bar\v=(\bar\omega(1-y/L)x_2+(\bar\omega/L)(R^2/3-x_2^2))\e_3.
\eqno(5.9)
$$
To fix the ideas, consider an axisymmetric particle whose shape in the co-rotating frame 
$\{x'_1,x'_2,x'_3\}$ $=\{x_1,x_2\cos\theta+x_3\sin\theta,x_3\cos\theta-x_2\sin\theta\}$ is:
$$
\Delta X(\hat\x)/R=\epsilon(-P_2(\hat x'_3)+\alpha P_3(\hat x'_3))
\eqno(5.10)
$$
with $\epsilon$ small, $\epsilon,\alpha>0$, and $P_2(z)=(3/2)z^2-1/2$ and $P_3(z)=(5/2)z^3-(3/2)z$ 
Legendre polynomials. Thus, for $\alpha=0$ the particle is an oblate ellipsoid, and, for
$\alpha>0$, the tip of the particle is at $x'_3>0$. (By composition of the harmonics in $\bar\v$ 
and in $\Delta X$, only $l=2$ and $l=3$ terms in $\Delta X$ give non-zero contribution to the 
matrix elements for force, torque and wall produced lift).

The particle instantaneous transversal velocity is the sum of the contribution from the wall, 
given by Eqn. (5.8), and the one due to interaction between the fore-aft asymmetry of the particle 
shape and the shear: $\dot y=v_\smalL+\frac{F_2}{6\pi\mu R}$. Substituting Eqns. (5.9-10) into Eqn. 
(5.8), and using Eqns. (3.8) and (5.6) gives $v_\smalL$. 
Repeating the same operation with $F_2$, with Eqns. (3.2) and (3.9) in place of (5.7-8), 
and using Eqn. (4.1):
$$
\dot y=\epsilon\bar\omega R\Big[(1-y/L)\Big(\frac{5}{7}\cos^2\theta-\frac{4}{7}\Big)\cos\theta
$$
$$
+\Big(\frac{3}{56}(1-y/L)\sin 2\theta
+\frac{25\alpha R}{28L}\Big(\frac{5}{4}\cos^2\theta-1\Big)\cos\theta\Big)
\Big(\frac{R^2}{y^2}-\frac{R^2}{(2L-y)^2}\Big)\Big].
\eqno(5.11)
$$
Details of the calculation are given in Appendix A; see in particular Eqns. (A14-15).
Notice that no contribution from the quadratic part of the shear 
is present in the term in the first line of the equation: the one which doe not come from 
interaction with the wall. Anyway, this seems to be fortuitous and non-zero contributions 
may show up at higher orders in the perturbative series in $\epsilon$. 

The contribution to torque from $\Delta X$ is obtained working with Eqns. (3.4) and (3.10)
and this leads to the equation of motion for the particle orientation, using also Eqn. (4.2):
$\dot\theta=\frac{\bar\omega}{2}(1-y/L)+\frac{T_1}{8\pi\mu R^3}$. Hence, from Eqn. (A16):
$$
\dot\theta=\bar\omega(1-y/L)\Big(\frac{1}{2}+\frac{3\epsilon}{4}\cos 2\theta\Big)
-\epsilon\alpha\bar\omega\frac{R}{L}\Big(\frac{15}{8}\cos^2\theta
-\frac{9}{14}\Big)\sin\theta
\eqno(5.12)
$$
Thus, close to the axis (i.e. when $|L-y|<\epsilon\alpha R$), a particle with fore-aft asymmetric
shape will be forced to get the orientation $\theta=0$. Under these conditions, the dominant part
of the drift is the interaction with the shear $[$first line of Eqn. (5.11)$]$, which pushes 
the particle towards $y=L$. Away from it, the standard Jeffery solution in the small eccentricity 
limit \cite{jeffery22}, corresponding to a periodic motion regime, is recovered, and, as discussed 
at the end of the previous sub-section, the mean transverse drift is equal to zero.  
\vskip 20pt

\centerline{\bf VI. Conclusions}
\vskip 5pt
The present work has been carried on along two parallel lines: the derivation of qualitative 
results on particle behaviors, by means of simple geometrical reasoning, 
and actual quantitative calculations, using perturbation theory around the spherical particle 
case. Emphasis was put, especially in the geometrical proofs, on the origin of the results in 
the symmetries of Stokes dynamics. It was thus possible to obtain a classification of the 
behaviors that a particle, depending on its shape, can have in linear and quadratic bounded 
shear flows. The present analysis was limited to 
shapes with at least two planes of symmetry, so that a large number of more irregular, 
strongly non-spherical objects were excluded. Also, some of the results, namely Eqn. (5.5) 
rested on a stability hypothesis for configurations, in which one of the particles symmetry planes
coincide with the shear plane. (It should be mentioned that recent results 
\cite{yarin97} suggest
instability of such configurations, and also chaos, at least in the case of very 
elongated ellipsoids).
In spite of these limitations, it was possible, in several important cases, to give a more direct 
physical 
content to results about the functional form of torques and forces, without actually having
to solve Stokes equation; one example is the generalization of Jeffery's result for the torque 
on an ellipsoid \cite{jeffery22}, provided by Fig. 4. 

One of the motivation for the present study was to get some ideas on the 
mechanisms determining transverse migration in plane shear flows. 
Some of the results obtained, were implicit in previous works on the behavior of droplets 
\cite{chan79} and vesicles \cite{helmy82,leyrat94} in viscous shear flow. However, an
explanation was missing, of which ingredients in the dynamics of these systems, were 
at the basis of the migration process. One result has been to clearly illustrate the
role of quadratic shear and of particle deformability. Summarizing, a rigid particle 
will drift to the center of a channel flow, only if fore-aft asymmetric
and sufficiently close to the channel axis, for quadratic shear to be dominant. A 
particle will drift to the channel axis irrespective of its initial 
position, only if deformable. This may happen, either because of deformations, produced by the
quadratic shear, which break fore-aft symmetry in a preferential direction \cite{olla99a},
or because of the presence of tank-treading motions \cite{olla97a,olla97b}), or because of a
combination of the two effects, as in the case of droplets \cite{leal80}.

The goal of this paper was, from one side, to 
understand the origin of particle behaviors in their shape symmetries and in 
those of the fluid flow; from the other, it was to furnish an efficient method for carrying 
on calculations in the perturbative quasi-spherical regime. Apart of the illustrative
examples considered in section IV, the technique allowed to obtain quantitative informations on 
the transverse drift of axi-symmetrical particles in channel flows. As clear already from the
geometrical considerations presented before, a particle drifts basically because of
three mechanisms. These are the interaction of the $l=2$ harmonics in the particle shape 
with the wall, mediated by the linear part of the shear; the direct interaction of the 
$l=3$ shape harmonics with the linear part of the shear (even in the absence of walls); the 
interaction of the $l=3$ shape harmonics with the walls, mediated by the quadratic part of the
shear. For a rigid particle in a channel flow, only the second contribution  
plays a role, given the absence of steady drifts close to the walls. For different reasons,
the same situation is met in the case of quasi-spherical vesicles in channel flows
\cite{olla99b}.

One purpose of the bra-ket formalism described in Eqns. (3.9-10) and (5.8) was to provide 
a viable tool in situations where a reciprocal theorem approach is too cumbersome. 
In the solid particle cases, considered in this paper, the amount of work needed in the
analysis, was clearly equivalent to what would have been necessary in a reciprocal 
theorem approach. Thus, the choice of the technique was mainly a matter of taste.
The situation becomes different if the fluid is strongly non-Newtonian, or in
general, if one has to deal with constitutive laws of an unusual form. 
In a separate paper, \cite{olla99b}, the present technique
was used to study the dynamics of inextensible membranes. In that case, the presence
of constitutive laws different from incompressibility made an approach based on ordinary
perturbation theory, rather than on the reciprocal theorem, the only reasonable choice.

\vskip 10pt
\noindent{\bf Aknowledgements}: I would like to thank Howard Stone and Dominique Barthes-Biesel for 
interesting and helpful conversation. Part of this research was carried on at CRS4 and at the 
Laboratoire de Mod\'elisation en M\'ecanique in Jussieu. I would like to thank Gianluigi Zanetti 
and Stephane Zaleski for hospitality.

\vskip 20pt

\centerline{\bf Appendix A. Calculation of vector spherical harmonics components}
\vskip 5pt
\renewcommand{\theequation}{\Alph{section}\arabic{equation}}
\setcounter{section}{1}
Matrix elements involving vector spherical harmonics, in the form given by Eqn. (2.3)
can be expressed in terms of integrals containing only plain spherical harmonics. For
instance, it is possible to write:
$$
v^{\rm e}_{lm}(R)=\frac{3}{R^3(l(l+1))^\frac{1}{2}}\int_0^Rx^2\d x\int\d\Omega_x
(x\nabla Y^*_{lm}(\x_i))\cdot\v(R\hat\x)
$$
$$
=(l(l+1))^{-\frac{1}{2}}\int\d\Omega_x Y_{lm}^*(\hxi)
         [3\hat\x\cdot \v(R\hat\x)-\nabla\cdot (x\v(R\hat\x))]_{x=R}
\eqno({\rm A}1)
$$
which is the second of Eqn. (2.4); one obtains the third equation in that system in completely 
analogous manner. For ease of notation, it is possible to introduce operators $\hat{\bf 0}^\mu_{lm}$
such that Eqn. (2.3) read:
$$
v^\mu_{lm}(R)=\langle lm|\hat{\bf 0}^\mu_{lm}\cdot\v\rangle
\eqno({\rm A}2)
$$ 
where $\langle lm|f\rangle=\int\d\Omega_xY^*_{lm}(\hxi)f(\x)$ indicates a plain spherical harmonics 
component.

For the analysis in section V, it is necessary to calculate quantities in the form 
$\hat{\bf 0}^\mu_{lm}\cdot\bar\v$ with $\bar\v=[0,0,x_2]$, $[0,0,1/3-x_2^2]$ and
$\hat{\bf 0}^\mu_{lm}\cdot\Delta X|Z\rangle$, where $Z=F_2,T_1,L$ and $\Delta X$ is 
given by Eqn. (5.10). Taking for simplicity $R=1$ from now on, one obtains promptly, 
in the linear shear case:
$$
\hat{\bf 0}_{lm}^{\rm e}\cdot\bar\v=\frac{3\hat x_2\hat x_3}{\sqrt{l(l+1)}}
\qquad
\hat{\bf 0}_{lm}^{\rm m}\cdot\bar\v=-\frac{\hat x_1}{\sqrt{l(l+1)}}
\eqno({\rm A}3)
$$
while, in the quadratic case:
$$
\hat{\bf 0}_{lm}^{\rm e}\cdot\bar\v=\frac{2\hat x_3(\hat x_1^2-5\hat x_2^2+\hat x_3^2)}
{3\sqrt{l(l+1)}}
\qquad
\hat{\bf 0}_{lm}^{\rm m}\cdot\bar\v=\frac{2\hat x_1\hat x_2}{\sqrt{l(l+1)}}
\eqno({\rm A}4)
$$
From here, the velocity components for the unperturbed flow are obtained. For the linear shear case:
$$
\bar U^{\rm s}_{2,\pm 1}={\rm i}\sqrt{\pi/15},
\qquad
\bar U^{\rm e}_{2,\pm 1}={\rm i}\sqrt{\pi/5}
\quad{\rm and}\quad
\bar U^{\rm m}_{1,\pm 1}=\pm\sqrt{\pi/3}
\eqno({\rm A}5)
$$
In the quadratic shear case:
$$
\bar U^{\rm s}_{10}=\frac{4\sqrt{3\pi}}{45}, 
\qquad
\bar U^{\rm e}_{10}=-\frac{2\sqrt{6\pi}}{45},
\qquad
\bar U^{\rm s}_{30}=\frac{2\sqrt{7\pi}}{35},
\qquad
\bar U^{\rm e}_{30}=\frac{4\sqrt{21\pi}}{105},
$$
$$
\bar U^{\rm s}_{3,\pm 2}=\sqrt{\frac{2\pi}{105}},
\qquad
\bar U^{\rm e}_{3,\pm 2}=\frac{2\sqrt{70\pi}}{105}
\quad{\rm and}\quad
\bar U^{\rm m}_{2,\pm 2}=\mp\frac{2{\rm i}\sqrt{5\pi}}{15}.
\eqno({\rm A}6)
$$
The calculation of the terms $\hat{\bf 0}^\mu_{lm}\cdot\Delta X|Z\rangle$, necessary for the
matrix elements $\langle Z|\Delta X|\mu lm\rangle$ entering Eqns. (3.9-10) and (5.8) is
more cumbersome. Life becomes simpler, however, exploiting the knowledge of the dependence 
on the orientation $\theta$ that the resulting forces, torques and drifts must have. 
To make an example, as discussed in section V, the contribution from $P_2(z')$ to $v_\smalL$, 
is necessarily proportional to $\sin 2\theta$. Isolating that contribution to $P_2(z')$,
i.e. $-(3/2)\hat x_2\hat x_3$, one obtains for instance:
$$
\frac{\hat{\bf 0}_{lm}^{\rm e}\cdot P_2|L\rangle}{\sin 2\theta}=
-\frac{45}{64\pi}\frac{\hat x_2\hat x_3(2\hat x_1^2-3\hat x_2^2+2\hat x_3^2)}{\sqrt{l(l+1)}}
\eqno({\rm A}7)
$$
plus terms that will give zero after integration in $\d\Omega_x$. The final expressions for
the non-zero matrix elements $\langle Z|P_\alpha|\mu lm\rangle$ are given below. For the lift 
from the wall:
$$
\langle L|P_2|{\rm e}2,\pm 1\rangle=
\frac{3{\rm i}}{224}\sqrt{\frac{5}{\pi}}\sin 2\theta,
\qquad
\langle L|P_2|{\rm m}1\pm 1\rangle=
\pm\frac{3}{32}\sqrt{\frac{3}{\pi}}\sin 2\theta
\eqno({\rm A}8)
$$
and
$$
\begin{array}{ll}
\langle L|P_3|{\rm e}10\rangle=
\frac{3}{112}\sqrt{\frac{6}{\pi}}(5\cos^2\theta-4)\cos\theta,
\qquad
\langle L|P_3|{\rm e}30\rangle=
\frac{1}{672}\sqrt{\frac{21}{\pi}}(5\cos^2\theta-4)\cos\theta,
\\
\langle L|P_3|{\rm e}3,\pm 2\rangle=
\frac{1}{448}\sqrt{\frac{70}{\pi}}(5\cos^2\theta-3)\cos\theta,
\ 
\langle L|P_3|{\rm m}2,\pm 2\rangle=
\mp\frac{3{\rm i}}{112}\sqrt{\frac{5}{\pi}}(4-5\cos^2\theta)\cos\theta.
\end{array}
\eqno({\rm A}9)
$$
For the force, indicating $|f\rangle=(6\pi\mu R)^{-1}|F_2\rangle$:
$$
\langle f|P_3|{\rm e}2,\pm 1\rangle=
\frac{\rm i}{70}\sqrt{\frac{5}{\pi}}(4-5\cos^2\theta)\cos\theta,
\eqno({\rm A}10)
$$
and
$$
\begin{array}{ll}
\langle f|P_2|{\rm e}10\rangle=
-\frac{1}{40}\sqrt{\frac{6}{\pi}}\sin2\theta,
\qquad
\langle f|P_2|{\rm e}30\rangle=
-\frac{1}{70}\sqrt{\frac{21}{\pi}}\sin 2\theta,
\\
\langle f|P_2|{\rm e}3,\pm 2\rangle=-\frac{\sin 2\theta}{\sqrt{70\pi}},
\qquad
\langle f|P_2|{\rm m}2,\pm 2\rangle=\pm\frac{\rm i}{20}\sqrt{\frac{5}{\pi}}\sin 2\theta.
\end{array}
\eqno({\rm A}11)
$$
Finally for the torque, indicating $|t\rangle=(8\pi\mu R^2)^{-1}|T_1\rangle$:
$$
\langle t|P_2|{\rm e}2,\pm 1\rangle=\mp\frac{3\rm i}{40}\sqrt{\frac{5}{\pi}}\cos 2\theta,
\qquad
\langle t|P_2|{\rm m}1,\pm 1\rangle=\frac{1}{40}\sqrt{\frac{3}{\pi}},
\eqno({\rm A}12)
$$
and
$$
\langle t|P_3|{\rm e}30\rangle=\frac{3}{112}\sqrt{\frac{21}{\pi}}(5\cos^2\theta-1)\sin\theta,
\quad
\langle t|P_3|{\rm e}3,\pm 2\rangle=\frac{3}{896}\sqrt{\frac{70}{\pi}}(4\cos^2\theta-1)\sin\theta
$$
$$
\langle t|P_3|{\rm m}2,\pm 2\rangle=\mp\frac{3\rm i}{140}\sqrt{\frac{5}{\pi}}\sin\theta.
\eqno({\rm A}13)
$$
The expressions for lift, force and torque are obtained substituting Eqn. (A5-6) into Eqn. (3.8)
and then, together with Eqns. (A8-13) into Eqns. (3.9-10) and (5.8). If no torques act on the 
particle, the particle will spin in such a way that $\hat U^{\rm m}_{1m}=\bar U^{\rm m}_{1m}$
in the third of Eqn. (3.8).

Using Eqns. (5.8) and (A8-9), one obtains for the wall contribution to lift, 
respectively, in the two cases 
$\Delta X=P_2(z')$, $\bar\v=[0,0,x_2]$, and $\Delta X=P_3(z')$, $\bar\v=[0,0,1/3-x_2^2]$:
$$
v_\smalL=-\frac{3\sin 2\theta}{56}f(L_l,L_r)
\quad{\rm and}\quad
v_\smalL=\frac{25}{56}\Big(\frac{5}{4}\cos^2-1\theta\Big)\cos\theta f(L_l,L_r)
\eqno({\rm A}14)
$$
where $f(x,y)=R^2(x^{-1}-y^{-2})$ $[$see Eqns. (5.1) or (B26)$]$.
Next, from Eqns. (3.9) and (A10-11), one finds for the drift produced by the force $F_2$ (i.e. the
contribution to drift that is independent from the wall), in the two cases $\Delta X=P_2(z')$, 
$\bar\v=[0,0,1/3-x_2^2]$ and $\Delta X=P_3(z')$, $\bar\v=[0,0,x_2]$:
$$
v_f=0
\quad{\rm and}\quad
v_f=\Big(\frac{5}{7}-\frac{4}{7}\cos^2\theta\Big)\cos\theta
\eqno({\rm A}15)
$$
where $v_f=(6\pi\mu R)^{-1}F_2$.  
Finally, from Eqns. (3.10) and (A12-13), one finds the $\O(\Delta X)$ correction to the
particle angular velocity $\dot\theta^\smalun$, in the two cases $\Delta X=P_2(z')$, 
$\bar\v=[0,0,x_2]$ and $\Delta X=P_3(z')$, $\bar\v=[0,0,1/3-x_2^2]$:
$$
\dot\theta^\smalun=-\frac{3}{4}\cos 2\theta
\quad{\rm and}\quad
\dot\theta^\smalun=\Big(\frac{9}{14}-\frac{15}{8}\cos^2\theta\Big)\sin\theta
\eqno({\rm A}16)
$$

\vskip 20pt
\centerline{\bf Appendix B. Wall contributions to transverse drift in channel flows}
\vskip 5pt
\setcounter{section}{2}
Consider a fluid flow, bounded by two walls perpendicular to the $x_2$ axis, 
located respectively at $x_2=L_r$ and $x_2=-L_l=L_r-2L$. 
We want to calculate the lift along the $x_2$ direction, on a small
particle placed at the origin of the axes, in the regime $R\ll L_l,L_r$, where $R$ is
the size of the particle. In this regime, if the the flow lines are taken parallel 
to the walls, the lift will be given by the image at the walls of the far field part
of the velocity perturbation. 

It is therefore necessary to know the velocity in the interval $-L_l<x_2<L_r$,
given boundary conditions at $x_2=L_r$, $x_2=-L_l$. It is expedient to Fourier transform in 
$x_{1,3}$. The vorticity and continuity equations, given stationary, creeping flow conditions
lead to the system:
$$
\begin{cases}
\partial_2v_3-ik_3v_2=a\exp(k(x_2-L_r))+b\exp(-k(x_2+L_l))\\
      i(k_3v_1-k_1v_3)=f\exp(k(x_2-L_r))+g\exp(-k(x+L_l))\\
      ik_1v_2-\partial_2v_1=c\exp(k(x_2-L_r))+d\exp(-k(x+L_l))\\
      \partial_2v_2+i(k_1v_1+k_3v_3)=0.\\
\end{cases}
\eqno({\rm B}1)
$$
where $k=\sqrt{k_1^2+k_3^2}$.  Multiplying the first and third 
equations in this system by $k_3$ and $k_1$ and taking the difference, one gets:
$$
\partial_2i(k_1v_1+k_3v_3)+k^2v_2=i(k_3a-k_1c)\exp(k(x_2-L_r))+i(k_3b-k_1d)\exp(-k(x_2+L_l))
\eqno({\rm B}2)
$$
and using the continuity equation:
$$
(\partial_2^2-k^2)v_2=-i(k_3a-k_1c)\exp(k(x_2-L_r))-i(k_3b-k_1d)\exp(-k(x_2+L_l))
\eqno({\rm B}3)
$$
which leads to the expression for $v_2$:
$$
v_2  = \Big(\hat v_2^r+iA(L_r-x_2)\Big)\exp(k(x_2-L_r))
     + \Big(\hat v_2^l+iB(L_l+x_2)\Big)\exp(-k(x_2+L_l))
\eqno({\rm B}4)
$$
where:
$$
A=\frac{k_3a-k_1c}{2k}\quad {\rm and}\quad B=\frac{k_3b-k_1d}{2k}
\eqno({\rm B}5)
$$
Substituting into the first and third of Eqn. (B1) and integrating in $x_2$: 
$$
\left\{\begin{array}{ll}
v_3 & = \hat v_3+\Big(\frac{ik_3\hat v_2^r+a}{k}-\frac{k_3A}{2k^2}
          +\frac{k_3A}{2k}(x_2-L_r)\Big)\exp(k(x_2-L_r))\\
       &-\Big(\frac{ik_3\hat v_2^l+b}{k}-\frac{k_3B}{2k^2}
          -\frac{k_3B}{2k}(x_2+L_l)\Big)\exp(-k(x_2+L_l))\\
v_1 & = \hat v_1+\Big(\frac{ik_1\hat v_2^r-c}{k}-\frac{k_1A}{2k^2}
          +\frac{k_1A}{2k}(x_2-L_r)\Big)\exp(k(x_2-L_r))\\
       &-\Big(\frac{ik_3\hat v_2^l-d}{k}-\frac{k_1B}{2k^2}
          -\frac{k_1B}{2k}(x_2+L_l)\Big)\exp(-k(x_2+L_l))
\end{array}
\right.
\eqno({\rm B}6)
$$
The integration constants $\hat v_{1,3}$ are eliminated using the second and fourth of Eqn. (B1).
Noticing the fact that the left hand side of those equations must be combinations of terms 
proportional to $\exp(\pm kx_2)$, we get:
$$
\begin{cases}
k_3\hat v_1-k_1\hat v_3=0\\
k_1\hat v_1+k_3\hat v_3=0\\
\end{cases}
\eqno({\rm B}7)
$$
leading to: $\hat v_{1,3}=0$. 

We now can enforce boundary conditions on $v_{1,3}$:
$$
\begin{cases}
v_3^r=\Big(\frac{ik_3\hat v_2^r+a}{k}-\frac{k_3A}{k^2}\Big)
+\Big(-\frac{ik_3\hat v_2^l+b}{k}+\frac{k_3B}{k^2}(1+2kL)\Big)\exp(-2kL)\\
v_3^l=-\Big(\frac{ik_3\hat v_2^l+b}{k}-\frac{k_3B}{k^2}\Big)
+\Big(-\frac{ik_3\hat v_2^l+a}{k}-\frac{k_3A}{k^2}(1+2kL)\Big)\exp(-2kL)\\
v_1^r=\Big(\frac{ik_1\hat v_2^r-c}{k}-\frac{k_1A}{k^2}\Big)
+\Big(-\frac{ik_1\hat v_2^l-d}{k}+\frac{k_1B}{k^2}(1+2kL)\Big)\exp(-2kL)\\
v_1^l=-\Big(\frac{ik_1\hat v_2^l-d}{k}-\frac{k_1B}{k^2}\Big)
+\Big(-\frac{ik_1\hat v_2^l-b}{k}-\frac{k_1A}{k^2}(1+2kL)\Big)\exp(-2kL)\\
\end{cases}
\eqno({\rm B}8)
$$
Adding the right and left terms, with factors $k_1$ and $k_3$ respectively for components
$1$ and $3$, we get:
$$
\begin{cases}
A+(2kL-1)\exp(-2kL)B=
-ik\hat v_2^r+ik\hat v_2^l\exp(-2kL)+k_3v_3^r+k_1v_1^r\\
B+(2kL-1)\exp(-2kL)A=
-ik\hat v_2^l+ik\hat v_2^r\exp(-2kL)-k_3v_3^l-k_1v_1^l\\
\end{cases}
\eqno({\rm B}9)
$$
Finally, the boundary conditions on $v_2$ are obtained from Eqn. (B4):
$$
\begin{cases}
v_2^r=\hat v_2^r+(\hat v_2^l+2iLB)\exp(-2kL)\\
       v_2^l=\hat v_2^l+(\hat v_2^r+2iLA)\exp(-2kL)\\
\end{cases}
\eqno({\rm B}10)
$$
At this point, the coefficients $A$, $B$ and $\hat v_2^{r,l}$ can be obtained from Eqns.
(B8-9):
\setcounter{equation}{10}
\begin{eqnarray}
\hat v^r_2 & = & C\Big(8ikL^2e^{-4kL}(k_1v_1^r+k_3v^r_3)+(1-(1-4kL+8(kL)^2)e^{-4kL})v_2^r 
\nonumber \\
           &   & +2iLe^{-2kL}(1-e^{-4kL})(k_1v_1^l+k_3v_3^l)
-(1+2kL+(2kL-1)e^{-4kL})e^{-2kL}v_2^l\Big)
\\
\hat v^l_2 & = & C\Big(-8ikL^2e^{-4kL}(k_1v_1^l+k_3v^l_3)+(1-(1-4kL+8(kL)^2)e^{-4kL})v_2^l
\nonumber \\
           &   & -2iLe^{-2kL}(1-e^{-4kL})(k_1v_1^r+k_3v_3^r)
-(1+2kL+(2kL-1)e^{-4kL})e^{-2kL}v_2^r\Big)
\end{eqnarray}
\begin{eqnarray}
A          & = & C\Big((1-(1+4kL)e^{-4kL})(k_1v_1^r+k_3v^r_3)+i((1-4kL)e^{-4kL}-1)kv^r_2
\nonumber \\
           &   & +((4kL-1+e^{-4kL})e^{-2kL}(k_1v^l_1+k_3v^l_3)+i(1+4kL-e^{-4kL})e^{-2kL}kv^l_2\Big)
\\
B          & = &
C\Big(-(1-(1+4kL)e^{-4kL})(k_1v_1^l+k_3v^l_1)+i((1-4kL)e^{-4kL}-1)kv^l_2
\nonumber\\
           &   & -((4kL-1+e^{-4kL})e^{-2kL}(k_1v^r_1+k_3v^r_3)+i(1+4kL-e^{-4kL})e^{-2kL}kv^r_2\Big)
\end{eqnarray}
where 
$$
C=(1-(2+16(kL)^2)e^{-4kL}+e^{-8kL})^{-1}
\eqno({\rm B}15)
$$
From Eqn. (B4), to calculate $v_2$ and therefore the lift, we do not need the coefficients $a-d$;
knowing $A$, $B$ and $\hat v_2^{r,l}$ is sufficient to the goal. Using Eqns. (B11-15),
we obtain:
$$
v_2=iF(L_r,k)(k_1v_1^r+k_3v_3^r)+G(L_r,k)v_2^r-iF(L_l,k)(k_1v_1^l+k_3v_3^l)+G(L_l,k)v_2^l
\eqno({\rm B}16)
$$
where:
$$
\begin{array}{ll}
F(L_y,k) & =  C(L_y-(L_y+8kL^2-4kLL_y)e^{-k(4L-2L_y)}\\
         & -(L_y-8kL^2+4kLL_y)e^{-4kL}+L_ye^{-k(8L-2L_y)})e^{-kL_y}
\end{array}
\eqno({\rm B}17)
$$
and
$$
\begin{array}{ll}
G(L_y,k) & = C(1+kL_y-(1+k(4L-L_y)+4k^2L(2L-L_y))e^{-k(4L-2L_y)}\\
         & -(1-k(4L-L_y)+4k^2L(2L-L_y))e^{-4kL}+(1-kL_y)e^{-k(8L-2L_y)})e^{-kL_y}
\end{array}
\eqno({\rm B}18)
$$
The lift is obtained evaluating the inverse Fourier transform of $v_2$ with respect to $x_1$ and 
$x_3$, at $\x=0$:
\setcounter{equation}{18}
\begin{eqnarray}
v_\smalL=-\frac{1}{(2\pi)^2}\int_0^\infty k\d k\int_0^\infty\hat x\d\hat x
\int_0^{2\pi}\d\psi\int_0^{2\pi}\d\phi\exp(-k\hat x\cos\psi)
\nonumber \\
\times\sum_{y=l,r}(\pm iF(L_y,k)(k_1v_1^y(\x)+k_3v_3^y(\x))+G(L_y,k)v_2^y(\x))
\end{eqnarray}
where $\hat x=\sqrt{x_1^2+x_3^2}$, $x_1=\hat x\cos(\psi+\phi)$ and $k_1=k\cos\phi$. The
$\pm$ signs go respectively with $y=l$ and $y=r$.
As boundary conditions on the wall, we have $\v^y=-\v$, where $\v$ is the far field part 
of velocity perturbation by the particle. In the absence of external forces, this is in 
the form:
\begin{equation}
\v(\x)=x^{-5}((\alpha-\beta)x_1^2+\beta x_2^2-\alpha x_3^2)\x
\end{equation}
plus terms which are not symmetric in $x_1$ and $x_3$, and give zero after integration.
Substituting into Eqn. (B19) and carrying out the integrals in $\phi$ and $\psi$, we get, 
in terms of Bessel functions $J_\mu(x)$:
\begin{equation}
v_\smalL=\frac{\beta}{2}\int_0^\infty k\d k\int_0^\infty\hat x\d\hat x
\sum_{y=l,r}\pm (G(L_y,k)L_yJ_0(k\hat x)
-F(L_y,k)k\hat xJ_1(k\hat x))\frac{2L_y^2-\hat x^2}{(\hat x^2+L_y^2)^{5/2}}
\end{equation}
where the following formulae for Bessel functions have been used \cite{abramowitz}:
\begin{equation}
\int_0^{2\pi}\d\psi\exp(-i\alpha\cos\psi)=2\pi J_0(\alpha)
\quad{\rm and}\quad
\int_0^{2\pi}\d\psi\cos\psi\exp(-i\alpha\cos\psi)=2\pi i J_1(\alpha)
\end{equation}
\begin{equation}
v_\smalL={\beta\over 4}(\frac{1}{L_l^2}-\frac{1}{L_r^2}).
\end{equation}
The $\hat x$-integration can be carried out analytically as well, using the formulae 
\cite{gradshteyn}:
$$
\int_0^\infty\frac{xJ_0(kx)\d x}{(x^2+L^2)^\frac{5}{2}}=\frac{1+kL}{3L^3}e^{-kL},
\qquad
\int_0^\infty\frac{x^3J_0(kx)\d x}{(x^2+L^2)^\frac{5}{2}}=\frac{2-kL}{3L}e^{-kL},
$$
\begin{equation}
\int_0^\infty\frac{x^2J_1(kx)\d x}{(x^2+L^2)^\frac{5}{2}}=\frac{ke^{-kL}}{3L}
\quad
{\rm and}
\quad
\int_0^\infty\frac{x^4J_1(kx)\d x}{(x^2+L^2)^\frac{5}{2}}=(1-\frac{kL}{3})e^{-kL}
\end{equation}
We thus get the final result:
\begin{equation}
v_\smalL=\frac{\beta}{2}\int_0^\infty k^2\d k\sum_{y=l,r}\pm(F(L_y,k)(1-kL_y)+G(L_y,k)L_y)e^{-kL_y}
\end{equation}
and, for $L_r\to\infty$, using the asymptotic expressions 
$F(L_l,k)=L_le^{-kL_l}$ and $G(L_l,k)=(1+kL_l)e^{-kL_l}$, this leads to the expression for
the lift obtained in \cite{olla97a}:
\begin{equation}
v_\smalL=\frac{\beta}{4L_l^2}
\end{equation}
$[$One factor $1/2$ was missing from \cite{olla97a}, Eqn. (51), due to a calculation mistake 
there$]$. In the general case, the remaining integral in $k$ must be carried out numerically. 
In Fig. 7 below, the dependence of $v_\smalL$ resulting from Eqn. 
(24) is compared with the analogous expression obtained using the large $L_y$ asymptotic 
expressions for $F$ and $G$: 
\begin{figure}[hbtp]\centering
\centerline{
\psfig{figure=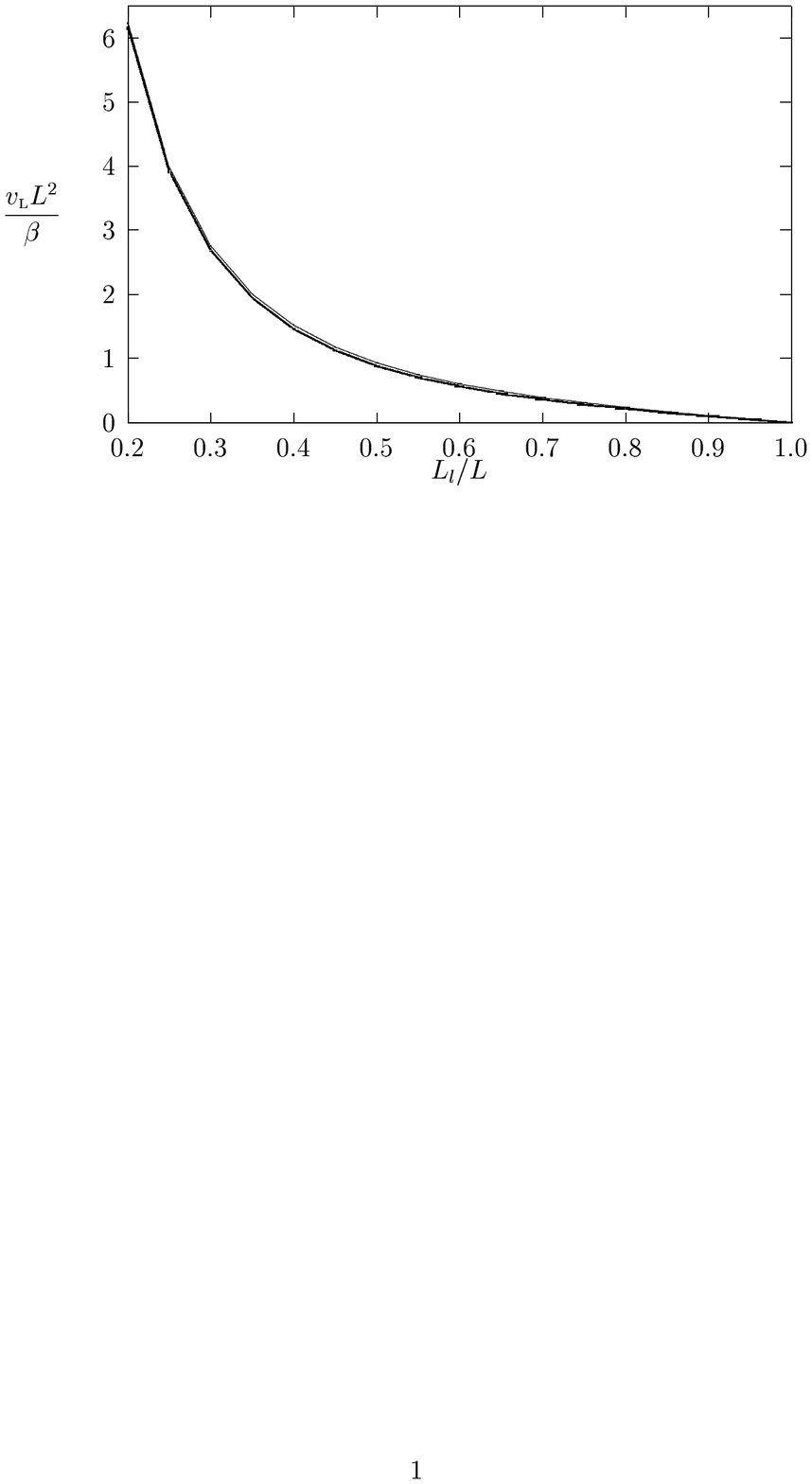,height=6.cm,angle=0.}
}
\caption{Behavior of the lift $v_\smalL$ in function of the distance $L_l$ of the particle from the
channel left wall. The results of Eqn. (B24) and of its approximated version Eqn. (B26) are 
almost undistinguishable}
\end{figure}

\end{document}